%% file: tau.tex
\documentclass[twoside,fleqn]{article}
\usepackage{espcrc2,epsfig,graphicx}
\newcommand{\E}[1]{$10^{#1}$eV}

\newcommand{\text}[1]{\mbox{#1}}

\title{Tau Neutrinos in the Auger Observatory~: A~New~Window~to~UHECR~Sources.}
\author{X. Bertou, P. Billoir, O. Deligny, C. Lachaud, A.  Letessier-Selvon
\address{L.P.N.H.E. Paris VI-VII, 4 place Jussieu, 75252 Paris, France} 
}
\begin{document}
\input {intro.tex}
\input{auger.tex}
\input{mc.tex}
\input{analytic.tex}
\input{resultats.tex}
\input{conclusions.tex}
\input{bib.tex}
\end{document}

%% file: intro.tex
\begin{abstract}
The cosmic ray spectrum has been shown to extend well beyond \E{20}. With nearly 20 events observed
in the last 40 years, it is now established that particles are accelerated or produced in the
universe with energies near \E{21} at the production site. In all production models neutrinos and photons are
part of the cosmic ray flux. In acceleration models (bottom-up models), they are produced as
secondaries of the possible interactions of the accelerated charged particle; in direct production models
(top-down models)  they are a dominant fraction of the decay chain.
In addition, hadrons above the GZK threshold energy will also produce,
along their path in the Universe, neutrinos and photons as secondaries of the pion photo-production
processes. Therefore, photons and neutrinos are very distinctive signatures of the
nature and distribution of the potential sources of ultra high energy cosmic rays.
In the following we describe the tau neutrino detection and identification capabilities 
of the Auger observatory.  We show that in the range $3\times 10^{17}-3\times 10^{20}$~eV 
the Auger effective apperture reaches a few tenths of
$km^2.sr$, making the observatory sensitive to fluxes as low as a few tau neutrinos
per $km^2.sr.year$. 
In the hypothesis of \mbox{$\nu_\mu\rightarrow\nu_\tau$} oscillations with full mixing, 
this sensitivity allows to a probe of the GZK cutoff as well as providing 
model independent constraints on the 
mechanisms of production of ultra high energy cosmic rays.
\vspace{1pc}
\end{abstract} 
\maketitle

\section{Introduction}
The origin of Ultra High Energy Cosmic Rays observed on Earth is a long lasting mystery\cite{Yoshida,Sigl,Bertou,Nagano}.
While the cosmic ray spectrum is now shown\cite{HiresTaup99,Agasa00} to extend beyond 
\E{20}, mechanisms producing or accelerating particles with energies near or above \E{21} are still uncertain.
\par
Only very powerful astrophysical objects can, in principle, produce these energies through conventional acceleration.
However the environment of the source itself generally prevents the accelerated particle to escape the site 
without severe energy losses, making such scenarios unlikely to explain the origin of UHECR. 
\par
Alternative hypotheses involving new physics such as collapse of Topological Defects (TD) or decay of Super Massive Relic Particles (SMRP)
are well suited to produce particles above \E{20} but they still lack a proof of existence. Moreover such models may reproduce 
the power law spectrum observed for the cosmic rays only at the condition that the decaying particle is much heavier than \E{20}.

\par
Transport from the source to Earth is also an issue. At those extreme energies the  Cosmic Microwave 
Background Radiation makes the Universe essentially opaque to protons, nuclei and photons which
suffer energy losses from pion photo-production, photo-disintegration or pair production. 
These processes led Greisen, Zatsepin and Kuzmin\cite{GZK} to predict a spectral cutoff 
around $5\times$\E{19}, the GZK cutoff. The available data, although still very scarce, do 
not support the existence of such a cutoff. Therefore the sources are either close by and locally
more dense for the cutoff not to show, or new physics modifies the expected energy 
losses of UHECR against the CMB photons.
\par
In this framework neutrino are an invaluable probe of the nature and the distribution of the potential sources. Essentially 
unaffected on their journey to Earth they may allow us to disentangle the source characteristics from the propagation distortions. 
In the following we will briefly describe the Auger observatory and show how $\nu_\tau$ are expected to interact and propagate in the 
Earth crust and be detected in Auger as low altitude and almost perfectly horizontal showers.  
In the framework of full $\nu_\mu\leftrightarrow\nu_\tau$ mixing we will
then evaluate our sensitivity to potential neutrinos sources and in particular to the low but almost certain flux of GZK neutrinos.

%% file: auger.tex
\section{Neutrino detection with the Auger detector}
Large area ground based detectors 
do not observe the incident cosmic rays directly but the Extensive Air Showers (EAS), a very large cascade of particles, that 
they generate in the atmosphere. All experiments aim to measure, as accurately as possible, 
the direction of the primary cosmic ray, its energy and its nature.
There are two major techniques used. One is to build a ground array of sensors spread over a large area,
to sample the EAS particle densities on the ground.
The other 
consists in studying the longitudinal development of the EAS by detecting
the fluorescence light emitted by the nitrogen molecules which are excited by the EAS secondaries.

The Auger Observatories\footnote{Named after the French physicist Pierre Auger.}~\cite{Auger} 
combine both techniques. The detector is designed to be fully
efficient for showers above 10~EeV \mbox{(1 EeV$\equiv10^{18}eV$)}, with a duty-cycle of
100\% for the ground array, and 10 to 15\% for the fluorescence
telescope.  The 1600 stations of the ground array are cylindrical \v{C}erenkov tanks
of 10~m$^2$ surface and 1.2~m height filled with filtered water; they are spaced 
by 1.5 km into a triangular grid. The construction started in the fall of 2000 in Argentina. 
Once completed in 2006, the observatories will be covering one site in each hemisphere. 
Their surface, 3000~km$^2$ each, will provide high statistics. With a total aperture of more than
14000~km$^2$sr, the Auger Observatories 
should detect every year of the order of 10000 events above 10~EeV and 100 above 100~EeV. 

Previous studies on UHE neutrino interaction in the atmosphere and
observation with Auger were reported in \cite{Zas2,Billoir}.
The idea of detecting $\nu_\tau$ interactions through the shower induced by the $\tau$ decays in
the atmosphere was presented in~\cite{Fargion2000,Puebla2000}.

In the following we will describe the specificity of the tau neutrinos
interaction and propagation, together with the possibility to detect the tau decay 
above the Auger ground array.

\subsection{Relevent properties of tau neutrinos}
Standard acceleration processes in astrophysical objects hardly produce
any $\nu_\tau$. In top-down models there is a full equivalence between all
flavors at the beginning of the decay chain but this symmetry breaks down at 
the end of the fragmentation process where the pions which
yield most of the expected neutrino flux are produced. 

\par 
This situation changes radically in the case of $\nu_\mu\leftrightarrow\nu_\tau$ oscillations with full mixing,
a hypothesis that seems to be supported by the atmospheric neutrino data and the K2K
experiment~\cite{Kamioka-nu2000}. In such a case the
$\nu_e:\nu_\mu:\nu_\tau$ flux ratios originally of $1:2:0$ evolves towards $1:1:1$
for a very wide range in $\delta m^2$ (given the very large distance
between the source and the Earth).  Half of the $\nu_\mu$ gets converted
into $\nu_\tau$ and all flavors are equally represented in the cosmic ray fluxes. 

\par
Unlike electrons which do not escape from 
the rocks or muons that do not produce any visible signal in the atmosphere\footnote{The electro-magnetic 
halo that surrounds very high energy muons does not spread enough in space to produce a detectable signal 
in an array of detectors separated by 1.5 km.}, taus, produced in the mountains or 
in the ground around the Auger array, can escape even from deep inside the rock and produce a clear
signal if they decay above the detector.

\par
The geometrical configuration that must be met to produce a visible
signal is rather severe. Neutrinos must be almost perfectly
horizontal (within  5 deg.). Therefore less than 10\% of the solid
angle is available while the neutrino energy and the distance between the interaction 
and the detector must match to have a good chance of observing the tau
decay. We will show in the following that indeed these criteria can be
met and that most of the
detectable signal (90\%) comes from upward going $\nu_\tau$ where the interactions
occur in the ground all around
the array and only 10\% from downward going $\nu_\tau$ coming from
interactions in the mountains surrounding the array.

\begin{figure}[!t]
\begin{center}
\hspace*{-0.50cm}
\includegraphics[width=8cm]{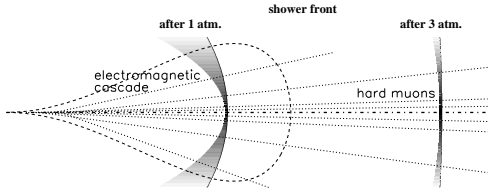}
\vspace*{-1.2cm}
\caption{Horizontal shower development.}
\label{showerdev}
\end{center}
\end{figure}
   
\subsection{Deep showers}
The UHE neutrinos may be detected and distinguished from ordinary hadrons by the 
shape of the horizontal EAS they produce.
Ordinary hadrons interact at the top of the atmosphere. At large
zenith angles (above 80~deg.) the distance from the shower maximum to the ground becomes larger
than 100~km.  At ground level the electromagnetic part of the shower is
totally extinguished (more than 6 equivalent vertical atmosphere were gone
through) and only high energy muon survive. In addition, the shower front is very flat (radius larger than 100~km) and the particles time
spread is very narrow (less than 50~ns).
\par
Unlike hadrons,  
neutrinos may interact deeply in the atmosphere and can initiate a shower in the volume of air immediately above the detector. This shower 
will appear as a ``normal'' one - although horizontal -, with a curved front
(radius of curvature of a few km), a large electromagnetic component, and
with particles well spread over time  
(over a few microseconds) [see Figure~\ref{showerdev}].
With such important differences, and if the fluxes are high enough,
neutrinos can be detected and identified. 
\begin{figure}[!t]
\includegraphics[width=8cm]{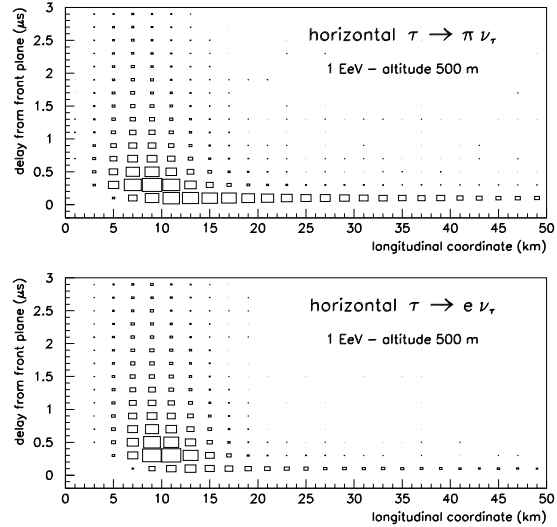}
\vspace*{-1cm}
\caption{Time structure of horizontal showers induced by a $\tau$ of 1 EeV.
The main difference is the importance of the muonic tail (concentrated at low delay times). 500m is the x altitude of the decay above the ground.}
\label{time_struct}
\end{figure}

\begin{figure*}[!t]
\begin{center}
\includegraphics[bbllx=0,bburx=525,bblly=340,bbury=640,clip=,width=16cm]{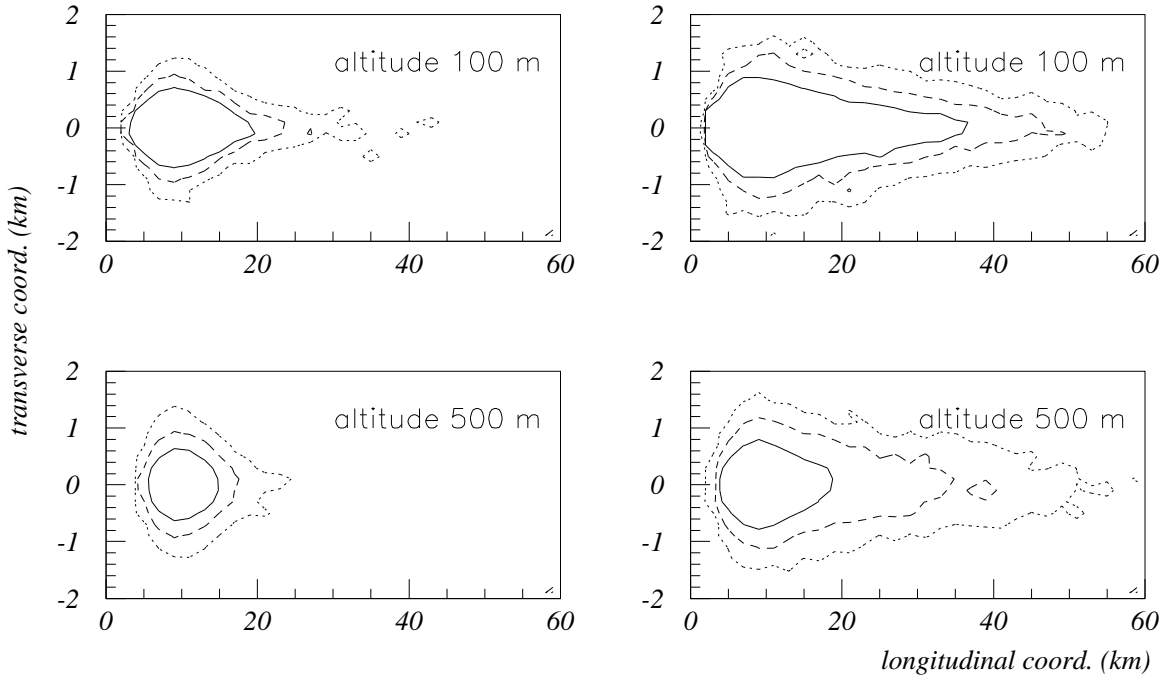}
\end{center}
\vspace*{-1cm}
\caption{Ground spots of horizontal showers induced by a $\tau$ of 1
EeV. Lines are iso-density curves at the threshold of the tank local trigger (solid), at 0.3 
(dashed) and at 0.1 (dotted). All of these data (even when below threshold) can be used from the global trigger
generated by a set of local triggers. Left: $\tau$ decay into $e \nu_e \nu_{\tau}$; 
right: decay into $\pi \nu_{\tau}$ }
\label{ground_spot}
\end{figure*}
  
\par
\label{shower_sim}
Showers produced from a $\tau$ decay have the same characteristics as
neutrino ones.
We simulated them with the AIRES program~\cite{AIRES}. A special mode, allowing the simultaneous
injection at a given point of several
particles with any direction was used. The development of the cascades in air,
with a thinning energy threshold of $E_{thin}=10^{-7}\,E_{\tau}$,
produced a set of weighted ``ground
particles'', which provides a good description of the densities expected
at each \v{C}erenkov tank.
It is important to mention that up going particles were allowed throughout
the cascade, as long as they remained inside a very large volume surrounding 
the shower maximum, while particles hitting the ground were not
followed further.     
\par
Fig.~\ref{time_struct} shows the time structure of two showers induced by 
horizontal tau decays at low altitude. 
The areas of the boxes
are proportional to the particle density and one can verify that the
shower maximum occurs around 10 km after the decay point and that the
``fat'' part of the shower extends over 10 to 15 km. As mentioned above, after 20
km or so, only muons well in time with the shower front survive. This is illustrated in the shower generated by a hadronic decay of the tau which contains more muons than a shower from an electronic decay 
as one can see in Fig.~\ref{time_struct} as the more important ``in time'' component at large
distance.

\par Horizontal showers, 
due to their {\em longitudinal} extension and provided that their core
is at sufficiently low altitude, may be seen in Auger at an energy much
lower than the vertical shower threshold of 10 EeV. 
For example, a horizontal shower induced by a primary particle of 0.1 EeV has an effective radius larger than 300 m 
over 10 km; if its core is at 100 m above the ground, it may easily
trigger 4 \v{C}erenkov tanks or more.
The extension of the shower core depends on the nature of the
primaries. Due to the large range of the muonic component, charged
hadrons give on average a larger footprint than  electrons or neutral pions.
\par
Fig.~\ref{ground_spot} shows examples of shower ground spots. The solid
contours give the area where the particle density is above the local
trigger threshold of the \v{C}erenkov tanks. On average, an electronic tau decay 
at 500 meters would trigger 3 stations, a pionic decay at 500 meters 5 
and at 100 meters more than 10.

%% file: mc.tex
\section{Tau events simulation}
\subsection{Interactions in earth}
\begin{figure}[!t]
\begin{center}
\includegraphics[bbllx=1,bburx=548,bblly=210,bbury=490,width=8cm,clip=]{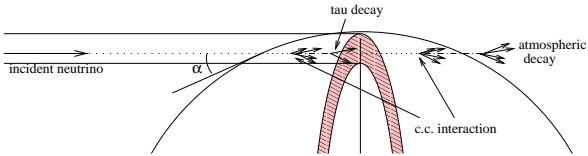}
\end{center}
\vspace*{-3.0cm}
\caption{Chain of interactions producing an observable shower.}
\vspace*{0.5cm}
\label{simul}
\end{figure}

A Monte-Carlo technique has been used to simulate the tau neutrino or charged lepton
interactions and propagation inside the Earth.
The lepton may interact several times through deep inelastic scattering, changing charge in most cases,
or eventually decay, but, in all cases, a tau neutrino or charged lepton is
present in the final state. Some energy is lost at each interaction, 
as well as continuously along the paths. However, in our energy range, the
initial direction of the incoming neutrino is always
conserved (Figure~\ref{simul}).
\par In our model, we assumed an isotropic incident flux of
neutrinos, an homogeneous Earth with density 2.15 g/cm$^3$~\cite{PDG}, and
a very high energy parameterization of the charged current (CC) cross section 
accurate to within 10\% as given by the results of the CTEQ4-DIS parton distributions~:
\[
\sigma_{\,cc}^{\,\nu N}=
1.0\,\left(\frac{E_\nu}{1\,EeV}\right)^{\!0.363}\!\!\!\!\!10^{-32}\mbox{\text cm}^2
\]
A step-by-step method is used; at each step, the probabilities of
different interactions and of the decay are evaluated as functions of the
energy. Both CC and NC interactions are taken into account, with
the cross section given above (and $\sigma_{\,nc}=0.4\,\sigma_{\,cc}$)~\cite{Gandhi1998} ;
the energy of the outgoing lepton is computed using a parametrization of the inelasticity at 1~EeV.

\par
Energy losses have been calculated including Bremsstrahlung (BS) and Pair 
Production (PP) as well as Deep Inelastic Scattering (DIS). 
The energy loss model is of the form~:
\[
-\frac{dE}{dx} = a + b(E)E
\]
where the second term is dominant above a few 100 GeV.
\par
Contributions from BS and PP have been rescaled from the muon values given 
in reference~\cite{PDG} and~\cite{Ginneken}, leading to $b = 0.08\times10^{-7}$ and 
$1.4\times10^{-7}$g$^{-1}$cm$^2$ respectively. We then obtained an 
attenuation length $L=(\rho\sum b)^{-1}$ of 31~km.
DIS contributions rely on parameterization of the photo-nucleon cross sections 
as well as on the proper modelisation of the nucleon structure functions at very low $x$ 
and/or very large $Q^2$. This subject being still tentative, we decided to use 
two different estimates,  an energy independant contribution 
(DIS-low, $b = 10^{-7}$g$^{-1}$cm$^2$) rescaled from the muon behavior given in~\cite{Ginneken} 
and an energy dependant one (DIS-high, $b = 6\,E_{18}^{0.2}\times 10^{-7}$g$^{-1}$cm$^2$ 
which dominates energy losses above $10^{15}$eV) as a parameterisation of the 
recent calculation from~\cite{Dutta}. This later case gives an attenuation length as low as
6~km at \E{18} strongly reducing the penetrating power of the tau.

 
 The $\tau$ is assumed to decay according to the relative
 probabilities into one of the most frequent modes~: $e$,
 $\mu$, $\pi$, $\pi\pi^0$, $\pi\pi^+\pi^-$ and $\pi\pi^0\pi^0$, which cover
 90\% of the total decays.
 We simplified
 the kinematical distribution of the decay products, reproducing only the essential feature, namely the
 fraction of
 initial energy going into the electromagnetic shower and into the
 hadronic one\footnote{Note that at 1~EeV the decay length of a $\pi^0$ is 
 200 m, therefore comparable to the interaction length in air
 (750m)}). A more accurate description would not modify our results as
 can be seen in Fig.~\ref{ground_spot} where the central ground region of
 a pure electromagnetic shower, although different, still compares with the
 hadronic one.
 The muons are considered to be unobservable. A possible effect of the
 longitudinal polarization of the tau was ignored.                                                           
 
\par Once a $\tau$ emerges from earth, and if a decay occurs within an
altitude of 3 km above the ground, an atmospheric shower is simulated as
described in Sec.~\ref{shower_sim}. 
The detector response is then evaluated through a 
simulation (outlined below) of the interactions of incident particles in water. 

\par Interactions in the mountains surrounding the detector were also simulated, using a detailed description of the relief. 
Their contribution was found to be much less than the material below sea level, whatever the energy. 
On the other hand we did not account for the lower density in Pacific Ocean, 250 km West from the southern site. 
The overall correction is less than 10 \%.

\subsection{Detector response}
The set of weighted ground particles in a ``sampling region'' around each
station is used to regenerate a set of particles entering the tank,
statistically reproducing all significant characteristics of the incident
flux~: global normalization of the different particles, distribution in
energy and direction.
\par
Then a simplified simulation is performed for interactions (cascade of
Compton scattering and pair production for photons, energy loss for
charged particles) and \v{C}erenkov emission in the water.
The production of \v{C}erenkov photons and their propagation in the tank
is performed until they hit a PMT or are absorbed in the water or in the tank walls.
The  PMT response is assumed to be proportional to the amount of light
emitted. This is a good approximation in most cases, in particular for the sum
over the three PMTs collecting the light from the tank.
\par The level of the local trigger (one tank) is set to 4 {\em vem}
(vertical equivalent muons), and a global trigger is built if at least 4
stations are locally triggered within 20 $\mu$s with a relatively compact
topology.  For exemple at least two stations must be within 3 km from a ``central'' one,
and an additional one within 6 km.
If needed, some long-shaped configurations with nearly aligned stations
and the right time spacing could be added to the global trigger
processor.  These allow to a gain of up to 50~\% in the acceptance at energies between 0.1
and 0.3 EeV. However, we did not include these additions in this study.
\begin{figure}[!t]
\begin{center}
\vspace*{-1cm}
\hspace*{-1cm} \includegraphics[width=8cm]{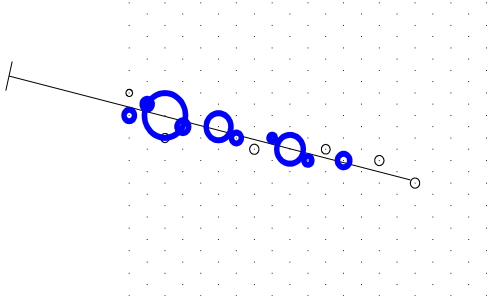}
\end{center}
\vspace*{-2cm}
\caption{Simulation of the ground trace left by a $\tau$ decay shower as
produced by a 5$\times$\E{17} tau neutrino.
Each thick circle represents a triggered station with a surface
proportional to the Cerenkov signal.
The $\tau$ shower had an energy of 3.63$\times$\E{17} and decayed 390
meters above the ground. In this particular exemple energy deposition in triggered tanks 
ranges from 4 to almost 100 {\em vem}.}\label{tau-events}
\end{figure}

\par  
Fig.~\ref{tau-events} shows a simulation
of the ground trace of a tau, produced by a $5\times$\E{17}
neutrino, as sampled by the Auger stations.
The signal is clearly visible and 10 stations pass the 4 {\em vem}
trigger requirement (thick circles).
 
\par
\begin{figure}[!ht]
\hspace*{-1.1cm}
\includegraphics[width=9.0cm]{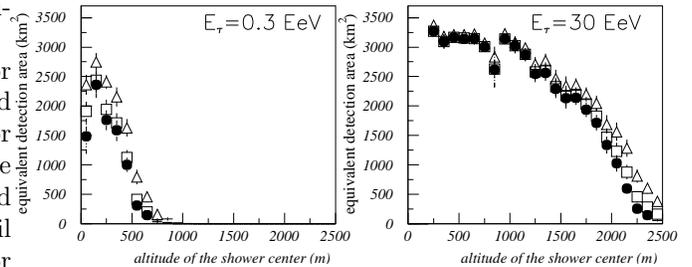}
\vspace*{-1.0cm}\caption{Equivalent detection area of tau showers averaged over all decay
channels versus the shower center (defined 10 km after the decay point)
altitude. Triangles : 3 stations global trigger, squares 4 stations, circles 4 stations
and the compact topology restriction.}
\label{detect}
\end{figure}

 
The probability to detect a shower with a given visible energy
depends essentially on the altitude of the core at the maximal lateral
development. It is not very sensitive to the exact definition of the local trigger
threshold nor to the global configuration.
For example, detecting all events with 3 locally triggered stations
would not increase sensibly the rate, except on the edges of the
acceptance below a few times 0.1 EeV. This is illustrated on Fig.\ref{detect},
where the equivalent area for detecting a shower has been plotted for various trigger
conditions~: triangles 3 stations, squares 4 stations and circles 4
stations plus the above condition on the global configuration. 

\subsection{Reconstruction}
The direction of origin may be estimated from the times of arrival of the
shower front on the stations, which is, as a first approximation, a plane
moving at speed $c$.
The precision on the azimuthal angle $\varphi$ is of the order of 1 deg,
and could be improved by taking into account the front curvature and by weighting
each station contribution according to its integrated amplitude.

\par
\begin{figure}[!ht]
\vspace*{-1.3cm}
\hspace*{-0.8cm}
\includegraphics[width=9cm]{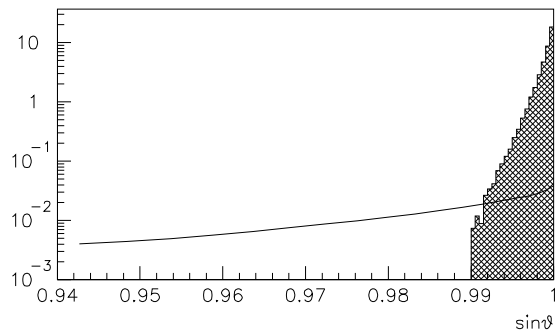}
\vspace*{-1.8cm}
\caption{$Sin\theta$ distribution of accepted events. Solid line~:
neutrino interaction in the atmosphere, filled area~: tau neutrino
interaction in the ground. The normalization between the two curves
(1:20) roughly corresponds to $\nu_\mu\leftrightarrow\nu_\tau$ oscillations with full
mixing and an $E^{-2}$ flux. The RMS on the reconstructed $sin\theta$ is at most 0.005.}
\label{sin_theta}
\end{figure}

 As a horizontal array is only sensitive to $\sin\theta$ the zenith angle 
$\theta$ is quite difficult to obtain precisely.
However, taus are all produced with $|\theta-90|<5$~deg. Therefore one can isolate them
from the standard horizontal neutrino shower as can be seen on Fig.~\ref{sin_theta}. 
\par 
The reconstruction of the energy $E_i$ of the incident neutrino is
 much more delicate~:
 \begin{itemize}
 \item The energy $E_{\tau}$ of the emerging tau may be much less
 than $E_i$, in particular for $E_i \gg 1$ EeV, where many intermediate
 interactions may have occurred reducing $E_{\tau}$ to a few 0.1 EeV.
 As $\theta$ is not well known, it is difficult to evaluate even an order
 of magnitude of the energy loss.
 \item An arbitrary fraction of $E_{\tau}$ goes into neutrinos
 and will not be visible while the decay type will influence
 the hadronic to electromagnetic ratio of the decay products. 
 This may be corrected for only if the tail of the shower is visible on
 the ground.
 \item The estimation of the shower energy depends strongly on the 
 altitude of the shower core which is {\em a priori} unknown. 
 If many stations are hit, there is a hope to evaluate it from 
 the transverse distribution.
 \end{itemize}
 Given these difficulties, we can predict the rate of events knowing 
 the energy spectrum of the original neutrinos. 
 The inverse will be difficult.
\par
A careful statistical analysis of all observable characteristics such as 
tank multiplicity, longitudinal and transverse profile of the ground spot and
time structure will certainly give additional information on the original spectrum.
We also beleive that for events where a arge number of tanks are struck we can obtain an estimate
of the neutrino energy but those studies need to be done.
 Of course, the hybrid reconstruction (involving both the ground array
 and the fluorescence detector of Auger) will be extremely valuable to
 remove some ambiguities (zenith angle, visible energy), 
 but such ``golden'' events are expected to be less than 10\% of the total
 event rate. 
  
\subsection{Evaluation of the acceptance}
The rate of observable events on a given surface $A$ (surface covered by the Ground Array) 
is simply the rate over the whole earth, multiplied by $A/(4\pi R_T^2)$ , where $R_T$ is 
the radius of the earth. This rate may be evaluated from a parallel flux
crossing the earth section ($\pi R_T^2$) as the integration over the
solid angle just gives an additional factor of $4\pi$. 
\par
A tau emerging with an angle $\alpha$ over the horizon greater than
$\alpha_m=0.3$~rad has no chance of producing an observable shower at
ground level. Therefore we only simulated incident neutrinos close to the earth
surface for which $\alpha \leq 0.3$~rad (hatched area on
Fig.\ref{simul}).
For various incident energies, $N_{sim} = 10^6$ neutrinos were simulated
and the complete history up to the trigger was followed, giving the total
number $N_{acc}$ of accepted events.
\par The apperture at a given energy may then be defined
as:\label{effective_area}
\begin{eqnarray*}
A_{\text{\small eff}} & = & 4\pi~\pi R_T^2 \sin^2\alpha_m~\frac{A}{4\pi
R_T^2}~\frac{N_{\text{\small acc}}}{N_{\text{\small sim}}}\\
& = & \pi A \sin^2\alpha_m~\frac{N_{\text{\small acc}}}{N_{\text{\small sim}}}
\end{eqnarray*}
and the rate of events (integrated over the solid angle) coming from neutrinos of energy between $E_1$ and
$E_2$ as~:
\[
\frac{dN_{\text{\small acc}}}{dt} = \int_{E_1}^{E_2} f(E)\,A_{\text{\small eff}}(E)\,dE
\]
where $f(E)$ is the incident flux.

%% file: analytic.tex
\subsection{Analytic estimation}
A simple analytic computation of the acceptance is easy if~:
\begin{enumerate}
\item the energy loss of tau leptons in the earth can be neglected;
\item only events with a unique interaction (single bang) in earth are considered;
\item the tau is assumed to carry  always the same fraction (80\%) of the
neutrino energy;
\item a simple geometric condition on the position of the tau decay can
model the detection condition.
\end{enumerate}
From our full  Monte-Carlo studies of the probability of detection of a tau shower
in Auger, given its energy, zenith angle, and altitude of decay
we found that a reasonable description of the largest detection altitude of the shower maximum was given
by~:
\[
h_m=1000+500\times\log(E_{18})
\]
where $h_m$ is the altitude in meters of the shower maximum.
\par
A more complete estimation can be done taking into account 
multiple interaction in earth (multi-bang events). 
Taking into account multi-bang events
increases considerably the acceptance at larger angles whatever the energy.
\par
\begin{figure}[!ht]
\begin{center}
\hspace*{-0.7cm}\includegraphics[width=8.cm]{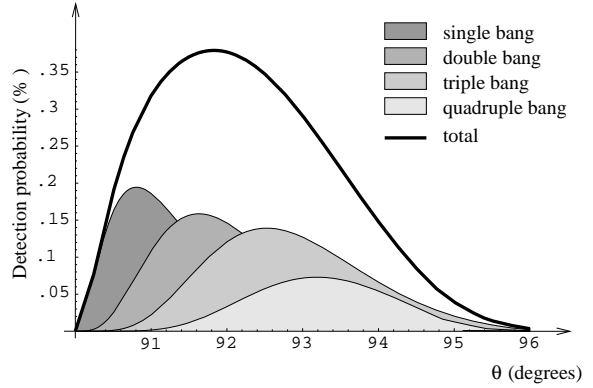}
\end{center}
\vspace*{-0.8cm}
\caption{Probability of detection of a tau shower for a 3.\E{18} neutrino
hitting the Auger array as a function of zenith angle.
}\label{fig:proba_math.eps}
\end{figure}

Fig.~\ref{fig:proba_math.eps}~shows the probability of producing and detecting a tau
shower as a function of its zenith angle. The contribution of the various multi-bang 
events can be seen.
\par
The acceptance can then be computed and compared with our full
Monte-Carlo results (see Fig~\ref{accept}). Differences come 
mainly from the approximations of the geometrical detection condition
and from the absence of energy loss in our analytical calculation.

%% file: resultats.tex
\section{Expected event rate}
\subsection{Acceptance}
\begin{figure}[!t]
\hspace*{-0.5cm}\includegraphics[width=7.7cm]{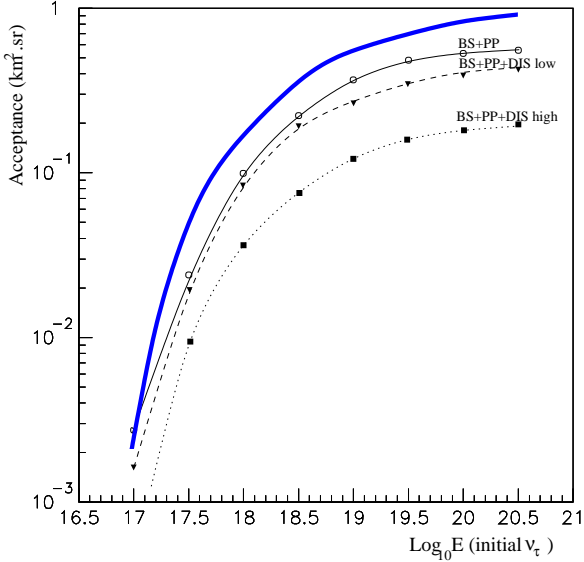}
\vspace*{-0.8cm}
\caption{Auger array effective aperture to $\nu_\tau$. Various continuous energy loss models
are presented. BS : bremmstrahlung, PP pair production, DIS deep inelastic scattering. The thick
solid line is our analytical calculation without energy loss.}
\label{accept}
\end{figure}

\begin{figure}[!b]
\hspace*{-0.3cm}\includegraphics[width=7.8cm]{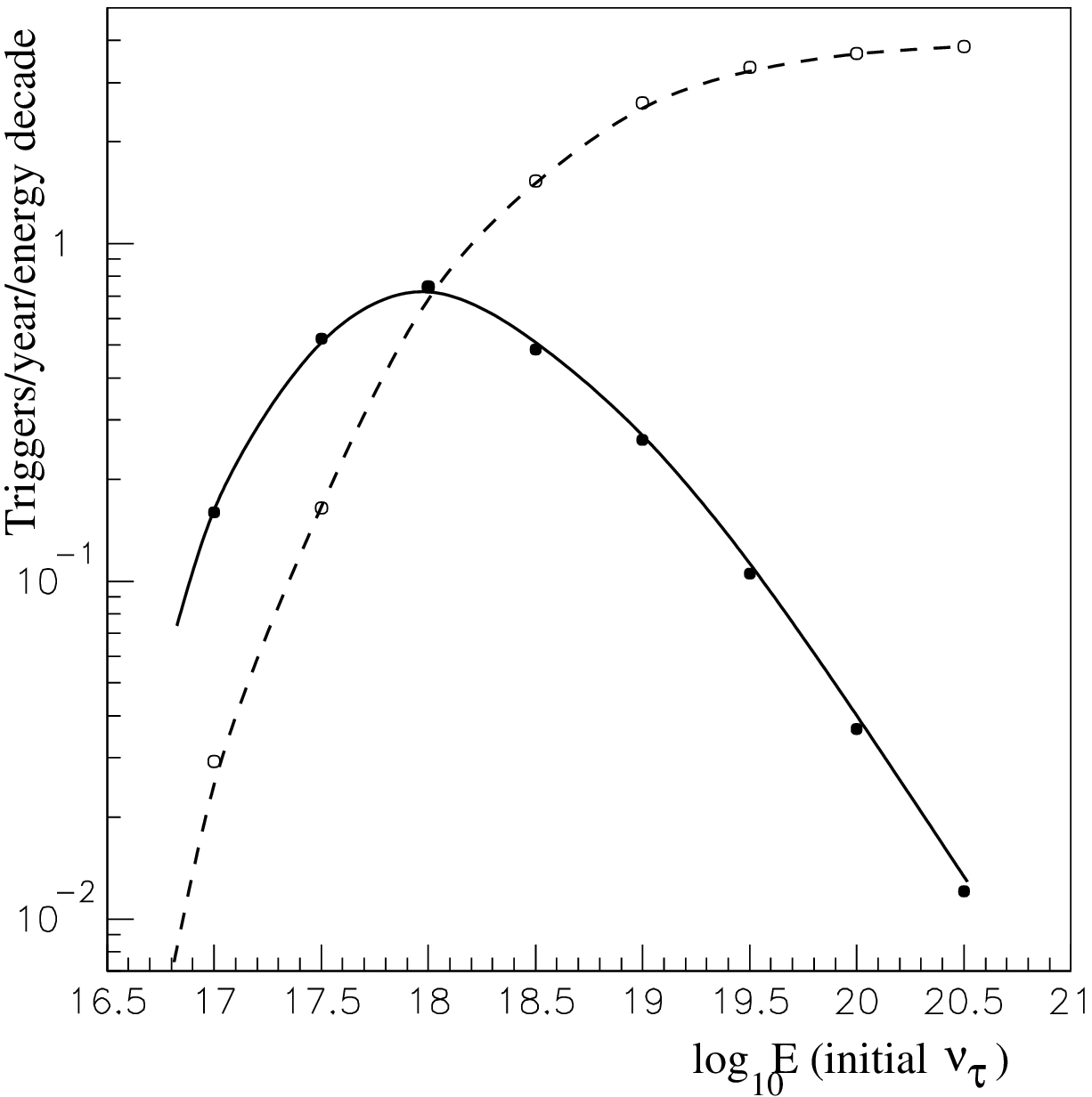}
\vspace*{-1.4cm}
\caption{The event rate per decade for BS+PP energy loss only.\newline
solid~line~$f(E)=3.1\,E_{18}^{-2}$~EeV$^{-1}$km$^{-2}$y$^{-1}$sr$^{-1}$; 
dashed~line~\mbox{$f(E)=3.1\,E_{18}^{-1}$~EeV$^{-1}$km$^{-2}$y$^{-1}$sr$^{-1}$.}
(3.1~EeV~km$^{-2}$y$^{-1}\equiv10^{-8}$~GeV~cm$^{-2}$s$^{-1}$).}
\label{accept2}
\end{figure}

The acceptance for $\nu_\tau$ interaction in the rocks surrounding
the Auger array is shown on Fig.\ref{accept}.
It may be represented either by the ``effective cross section'' defined
in Sec.~\ref{effective_area}, or, by introducing the event rate per
decade~:
\begin{eqnarray*}
I_{10} (E) & = & \lim_{\epsilon \rightarrow 0} 
\frac{\int_{E-\epsilon}^{E+\epsilon}f(e)A_{eff}(e)\,de } 
{\log_{10}{\left ( \frac{E+\epsilon}{E-\epsilon} \right )}}\\
&   = & \ln{10}\, E\, f(E) \, A_{eff}(E)
\end{eqnarray*}
With this definition, one can directly obtain the number of events detected
per year measuring the area under the curve $I_{10}(E)$ on a $I_{10}(E)$ versus log~E
plot as shown on Fig.~\ref{accept2}.
 
\par
\begin{figure}[!t]
\hspace*{-0.8cm}\includegraphics[width=8.5cm]{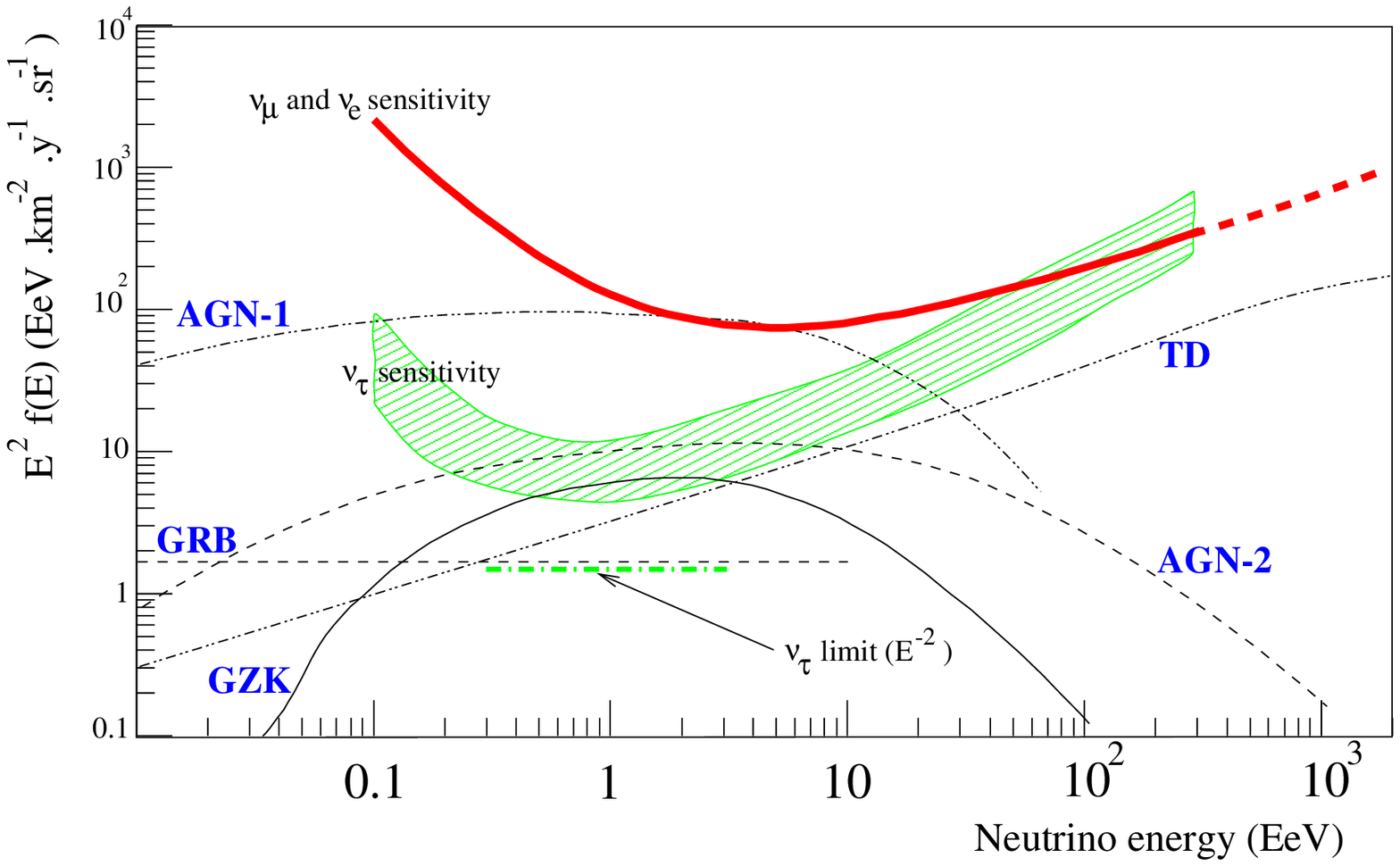}
\vspace*{-1.3cm}
\caption{Muon or tau neutrino and anti-neutrino fluxes
from various sources in the full mixing hypothesys, taken
from~\cite{Protheroe} (and divided by 2). Dotted lines are
speculative fluxes, dashed probable and solid certain.
The thick solid line and the hatched area represent the Auger sensitivity 
defined by $I_{10}(E)=1$, i.e. one event per year and per decade.
Top line for horizontal shower from $\nu_e$ and $\nu_\mu$ interactions in the atmosphere
hatched area for tau induced showers under strong DIS loss (top of the area) or no DIS loss (bottom). 
Any flux lying above those curves for at least one decade will give
more than one events per year in Auger.
We also plotted the 90\% C.L. limit (background free detection) for an E$^{-2}$ flux between 0.3 and 3 EeV
that Auger could acheive after five years.}
\label{nu-fluxes}
\end{figure}

We defined the Auger flux sensitivity as the neutrino flux giving at least one
observed event per decade of energy every year i.e. for which $I_{10}(E)=1$. 
This sensitivity is shown on Fig.~\ref{nu-fluxes}
together with the expected fluxes from a model calculation by Protheroe~\cite{Protheroe}.   
All predicted fluxes are $\nu_\mu$ fluxes, in the full mixing hypothesis
$\nu_\tau$ fluxes are half of those. 
The limits from atmospheric neutrino interaction formerly calculated in~\cite{Billoir} and from 
our present calculation of $\nu_\tau$ interactions in 
the rocks are shown.

\par 
For standard neutrino interactions in the atmosphere, each site of the Auger observatory reaches 
10 km$^3$ water equivalent (w.e.) of target mass at 1 EeV, and
only the models classified as speculative by Protheroe~\cite{Protheroe} are expected to yield a detectable signal.
However, for tau induced showers the target mass is increased by a factor
of about 30 at 1~EeV, allowing for a detectable signal even for the lowest
expected fluxes. The expected number of events per year from various
UHECR production models and from the GZK\footnote{One should note that the GZK neutrino model that we have
taken from~\cite{Protheroe} is almost one order of magnitude lower than
the prediction given in~\cite{Hill-Schramm} therefore we feel quite
confident that the Auger observatory  will observe a few
$\nu_\tau$ candidates if the $\nu_\mu\leftrightarrow\nu_\tau$ oscillation results are confirmed.}
neutrinos (a very low but almost
certain flux) are presented in Table~\ref{expected_events}. 
\begin{table}[!ht]
\caption{Expected number of events per year for the source models presented
in Fig.~\ref{nu-fluxes} and various DIS contributions to continuous energy losses.}\vspace*{0.2cm}
\begin{center}
\begin{tabular}{l|c|c|c|c|c}
DIS & AGN-1 & TD & GRB & GZK & AGN-2 \\ \hline
none & 27.0 & 2.3 & 0.5 & 1.7 & 2.9 \\ \hline
low  & 24.0  & 1.8 & 0.4 & 1.5 & 2.5 \\ \hline
high & 10.0 & 0.8  & 0.2 & 0.6 & 1.1
\end{tabular}\vspace*{-0.2cm}
\end{center}
\label{expected_events}
\end{table}
\par
The data in the table demonstrate the capability of the Auger detector to probe the
GZK neutrino flux. This is a crucial test as most acceleration
mechanisms of protons in cosmologically distributed sources as 
well as top-down models will produce a neutrino
flux at least equal to this one.                                                                             

%% file: conclusions.tex
\section{Conclusions}
The Auger observatories are found to have an optimal geometry for the detection of $\nu_\tau$
interaction in earth in the $0.1-100$~EeV energy range (the GZK range).
Indeed, above \E{17} the earth is not transparent to neutrinos and for 
tau or muon neutrinos successive charged/neutral current 
interactions will degrade the energy below \E{16}, mixing the high energy signal 
with the more standard and far more numerous low energy one. 
Therefore a maximum of a few 100 km of rocks should intersect the neutrino trajectory to limit the number 
of interactions and allow a high energy lepton (above \E{17}) to escape. 
Only nearly horizontal neutrinos interacting in 
mountains or in the top few kilometers of the Earth fulfill this
requirement. 
\par
At a few tens of EeV, the tau decay length is over 1000 km and the probability of a 
decay inside the field of view of current or foreseen detectors becomes very small.
Therefore, only the energy range $0.1-10$~EeV is truly visible. 
At these relatively ``low'' energies the fluorescence signal 
is rather small and will be difficult to see from far away (a few tens of km), a necessary condition to have
enough acceptance for observing a few events a year.

%% file: bib.tex
\font\nineit=cmti9
\font\ninebf=cmbx9
